\documentclass[twocolumn,showpacs,amsmath,amssymb,prl]{revtex4}
\usepackage{graphicx}
\usepackage{bm}
\usepackage{dcolumn}

\begin{document}
\title{Curvature-induced geometric potential in strain-driven nanostructures}

\author{Carmine Ortix}
\affiliation{IFW Dresden, Helmholtzstr. 20, D-01069 Dresden, Germany}

\author{Suwit Kiravittaya}  
\affiliation{IFW Dresden, Helmholtzstr. 20, D-01069 Dresden, Germany}

\author{Oliver G. Schmidt}
\affiliation{IFW Dresden, Helmholtzstr. 20, D-01069 Dresden, Germany}

\author{Jeroen van den Brink}
\affiliation{IFW Dresden, Helmholtzstr. 20, D-01069 Dresden, Germany}

\date{\today}

\begin{abstract}
We derive the effective dimensionally reduced Sch\"odinger equation for electrons in strain-driven curved nanostructures by adiabatic separation of fast and slow quantum degrees of freedom. The emergent strain-induced geometric potential strongly renormalizes the purely quantum curvature-induced potential and enhances the effects of curvature by several orders of magnitude. Applying this analysis to nanocorrugated films shows that this curvature-induced potential leads to strongly enhanced electron localization and the opening of substantial band gaps.
\end{abstract}

\maketitle

The experimental progress in synthesizing low-dimensional  nanostructures with curved geometries -- the next generation nanodevices \cite{ahn06ko10,mei07,par10} -- has triggered the interest in the theory of quantum physics on bent two-dimensional manifolds. The current theoretical paradigm relies on a thin-wall quantization method originally introduced by Da Costa \cite{dac82}. It treats the quantum motion on a curved two-dimensional (2D) surface as the limiting case of a particle in three-dimensional (3D) space subject to lateral quantum confinement. With Da Costa's method the surface curvature is eliminated from the Schr\"odinger equation at the expense of adding a potential term to it. This simplifies the problem substantially, as electrons now effectively live in 2D 
space, in presence of a curvature-induced potential. 

As the curvature induced potential is entirely of quantum origin -- it is in magnitude proportional to $\hbar$ -- its physical consequences in condensed matter systems can only be observed on the nano-scale. In this realm the curvature-induced geometric potential can cause intriguing phenomena, such as winding-generated bound states in rolled-up nanotubes \cite{ved00,ort10} and topological bandgaps in periodic curved surfaces \cite{aok01}. However, the experimental realization and exploitation of such phenomena is hindered by the fact that in actual systems with curvature radii of hundreds of nanometers, Da Costa's quantum potential is still very weak and typically only comes into play on the sub-kelvin  energy scale.

In this Letter, we mitigate this problem by developing a thin-wall quantization procedure which explicitly accounts for the effect of the deformation potentials of the model-solid theory \cite{van89,sun10}. By employing a method of adiabatic separation of fast and slow quantum degrees of freedom, we show that local variations of the strain render an emergent geometric potential that strongly (often gigantically) boosts the purely quantum geometrical potential.
%
%
%
 The theoretical framework that we develop is immediately relevant for electronic nanodevices as the present-day nanostructuring method \cite{ahn06ko10} is based on the tendency displayed by thin films detached from their substrates to assume a shape yielding the lowest possible elastic energy. As a result, thin films can either roll up into tubes \cite{sch01,pri00} or undergo wrinkling to form nanocorrugated structures \cite{fed06,mei07,cen09}. A key property of such bent nanostructures is the nanoscale variation of the strain. It for instance leads to considerable band-edge shifts \cite{den10} with regions under tensile and compressive strain shifting in opposite direction. To include such a nanoscale sequence of potential wells in the elastically relaxed structure one obviously  needs to go beyond Da Costa's thin-wall quantization framework since it intrinsically couples the transversal quantum degrees of freedom to the tangential quantum motion along the curved surface.

In the thin-wall approach quantum excitation energies in the normal direction are raised far beyond those in the tangential direction by the lateral confinement. Hence one can safely neglect the quantum motion in the normal direction thereby reaching an effective dimensionally reduced Schr\"odinger equation. As opposed to a classical particle, a quantum particle constrained to a curved surface retains some knowledge of the surrounding  3D space. In spite of the absence of interactions, it indeed experiences an attractive potential of geometrical nature \cite{dac82}.   It has been shown that Da Costa's thin-wall quantization procedure is well-founded, also in presence of externally applied electric and magnetic fields \cite{fer08ort11}.  Even more the experimental realization of  an optical analogue of the curvature-induced geometric potential \cite{sza10} has provided empirical evidence for the validity of this approach. 

\begin{figure}[tbp]
\includegraphics[width=.9\columnwidth]{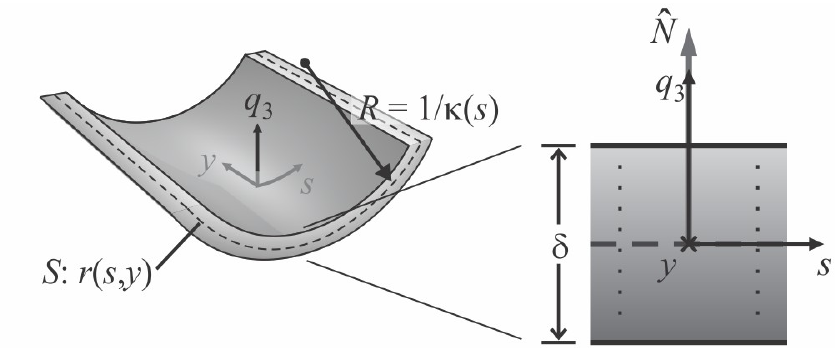}
\caption{Schematics of the three-dimensional coordinate system for a strain-driven bent nanostructure with positive radius of curvature $R$ and total thickness $\delta$. The $q_3=0$ surface ${\cal S}$ corresponds to the stress-free surface. Regions with $q_3>0$ ($q_3<0$) are under compressive (tensile) strain.}
\label{fig:schematic}
\end{figure}

To develop a theoretical framework for strain-induced geometric potentials we therefore use the same conceptual framework. We start the mathematical description by defining a 3D curvilinear coordinate system [see Fig.~\ref{fig:schematic}] 
for a generic bent nanostructure. The stress-free surface ${\cal S}$ is parametrized as ${\bf r}={\bf r}(s,y)$ where $y$ is the coordinate along the translational invariant direction of the thin film and $s$ is the arclength along the curved direction of the surface ${\cal S}$ measured from an arbitrary reference point. Nothing prevents the thin film to be deflected also along the $y$ direction. However, in the remainder we will neglect this effect for simplicity. The 3D portion of space of the thin film can be then parametrized as ${\bf R}(s,y,q_3)={\bf r}(s,y)+q_3 \hat{N}(s)$ with $\hat{N}(s)$ the unit vector normal to ${\cal S}$. We can evaluate the strain distribution in the thin film by assuming that along the undeflected direction $\epsilon_y \equiv 0$. It is well known \cite{lan86}
that the strain in the direction along the surface varies linearly across the thin film as $\epsilon_s = - q_3 \kappa(s)$ where $\kappa(s)$ is the principal curvature of the surface ${\cal S}$. The strain in the normal direction can be related to $\epsilon_s$ by means of the Poisson relation $\epsilon_{q_3}=-\left[\nu / \left(1-\nu\right) \right] \, \epsilon_s$ 
with $\nu$  the Poisson ratio. From the linear deformation potential theory \cite{van89}  we then have that the strain induced shift of the conduction band corresponds to a local potential for the conducting electrons ${\cal V}_{\epsilon}(s,q_3)= \gamma \, q_3 \kappa(s)$ with $\gamma > 0$ yielding an attraction toward regions under tensile strain [c.f. Fig.~\ref{fig:schematic}]. The characteristic energy scale $\gamma$ which can be explicitly computed in the different conduction valleys of the nanostructure  is proportional to the shear and the hydrostatic deformation potentials and typically lies in the eV scale for conventional semiconductors \cite{van89}. By  adopting Einstein summation convention, the Schr\"odinger equation for the quantum carriers in the effective mass approximation then takes the following compact form \cite{fer08ort11}
\begin{equation}
-\dfrac{\hbar^2}{2 m^{\star}} G^{i j} \,{\cal D}_i\, {\cal D}_j \,\psi + {\cal V}_{\epsilon}(s,q_3) \,\psi = E\, \psi.
\label{eq:startingequation}
\end{equation}
In the equation above $G^{i j}$ corresponds to the 3D metric tensor of our coordinate system, the covariant derivative ${\cal D}_i$ is defined as ${\cal D}_i= \partial_i v_j - \Gamma_{i j}^k v_k$ with $v_j$  the covariant components of a generic 3D vector field, and the $\Gamma_{i j}^k$'s are the Christoffel symbols. 
By expanding Eq.~(\ref{eq:startingequation}) by covariant calculus \cite{fer08ort11} we get 
\begin{eqnarray}
E \, \psi&=&\left[-\dfrac{\hbar^2}{2 m^{\star} \, H_{S}}  \partial_s \left(\dfrac{1}{H_S} \partial_s\right)-\dfrac{\hbar^2}{2 m^{\star} \, H_{S}} \partial_{q_{3}} \left(H_S \, \partial_{q_{3}}\right) \right. \nonumber \\ & & \left.-\dfrac{\hbar^2}{2 m^{\star}} \partial^2_y+ {\cal V}_{\epsilon}(s,q_3)+ {\cal V}_\lambda(q_3) \right] \psi, 
\label{eq:schro} 
\end{eqnarray}
where we defined $H_S=1 -\kappa(s) \, q_3$ , and we introduced a squeezing potential in the normal direction ${\cal V}_\lambda(q_3)$. In the following it will be considered as given by two infinite step potential barriers at $\pm \delta / 2 $ where $\delta$  is the total thickness of the thin film. However,  our results can be straightforwardly generalized to other types of squeezing potential, e.g. harmonic traps.  

\begin{figure}[tbp]
\includegraphics[width=.9\columnwidth]{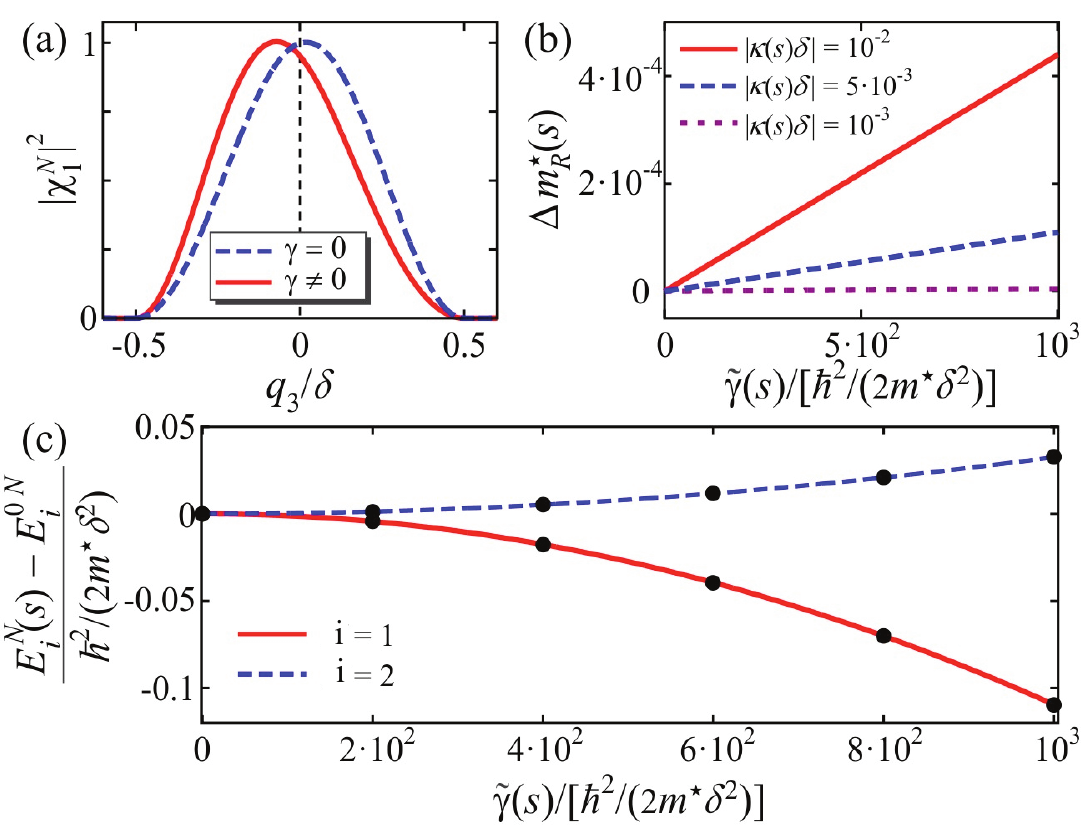}
\caption{(a) Schematics of the electron density probability in the lowest normal eigenstate as a function of the coordinate $q_3$ measured in units of the thin film thickness $\delta$ for a positive value of the local curvature in a strain-free ($\gamma=0$) and a strained ($\gamma \neq 0$) curved nanostructure.  
The vertical dashed line indicates the stress-free surface $q_3=0$.  (b) Deviation of the locally renormalized effective mass $\Delta m^{\star}_R(s)=(m^{\star}_R(s) -m^{\star}) / m^{\star}$  as a function of the typical energy scale $\widetilde{\gamma}(s)$  for different values of the expansion parameter $\kappa(s) \delta$ . (c) Behavior of the adiabatic potentials $E^N(s)$ for the first two transversal subband measured from the quantum well levels $\hbar^2 \pi^2 i^2 / (2 m^{\star} \delta^2)$ for $\kappa(s) \delta = 10^{-2}$  as a function of  $\widetilde{\gamma}(s)$. The continuous lines indicate the exact behavior whereas the points are the results of the perturbation theory approximation.}
\label{fig:normalwaves}
\end{figure}

In the same spirit of the thin-wall quantization procedure \cite{dac82} we next introduce a new wavefunction $\chi$ for which the surface density probability is defined as $\int |\chi(s,y,q_3)|^2 d q_3$. Conservation of the norm requires $\chi=\psi \times H_S^{1/2}$. The resulting Schr\"odinger equation is then determined by the Hamiltonian 
\begin{eqnarray}
{\cal H}&=&-\dfrac{\hbar^2}{2 m^{\star}} \partial_y^2  - \dfrac{\hbar^2}{2 m^{\star}} \dfrac{\partial_s^2 }{H_S^2}+\dfrac{\hbar^2}{m^{\star}} \dfrac{\partial_s H_S \, \partial_s}{H_S^3} -\dfrac{\hbar^2}{2 m^{\star}} \partial_{q_{3}}^2 \nonumber \\ & & 
-\dfrac{\hbar^2}{2 m^{\star}} \, \left[\dfrac{5}{4} \, \dfrac{\left(\partial_s H_S \right)^2}{H_S^4}-\dfrac{\partial_s^2 H_S}{2 \, H_S^3} \right]-\dfrac{\hbar^2}{2 m^{\star}} \dfrac{\left(\partial_{q_{3}} H_S \right)^2}{4 \, H_S^2} \nonumber \\ & &  +  {\cal V}_{\lambda} (q_3) + {\cal V}_{\epsilon}(s,q_3). \label{eq:effectivehamiltonian}
\end{eqnarray}
Assuming the thickness of the thin film $\delta$ to be small compared to the local radius of curvature $R(s)=\kappa(s)^{-1}$, the Hamiltonian Eq.~(\ref{eq:effectivehamiltonian}) can be expanded as  ${\cal H}=\sum_k q_3^k \, {\cal H}_k$. 
At the zeroth order in $q_3$ we recover precisely the effective Hamiltonian originally introduced by Da Costa \cite{dac82} which disregards the strain-induced shifts of the conduction band and guarantees the separability of the tangential motion from the transverse one. In this case the effect of the curvature eventually results in the well-known geometric potential ${\cal V}_{g}(s)=-\hbar^2 \kappa(s)^2 / (8 \, m^{\star})$. 
Strain effects can be explicitly monitored by retaining linear terms in $q_3$ in which case we obtain the effective Hamiltonian
\begin{eqnarray}
\widetilde{{\cal H}}&=&-\dfrac{\hbar^2}{2 m^{\star}} \left(\partial^2_y+ \partial_s^2+\partial_{q_{3}}^2\right)+ {\cal V}_g(s)+ {\cal V}_{\lambda}(q_3) +  {\cal V}_{\epsilon}(s,q_3) \nonumber \\ &- & \dfrac{\hbar^2}{4 m^{\star}}  \left[\partial_s^2 \kappa(s) +  \kappa(s)^3 \right]  q_3  -\dfrac{\hbar^2}{m^{\star}} q_3 \,\partial_s \left[\kappa(s) \partial_s \right] .\label{eq:effectivehamiltonianapprox}
\end{eqnarray}
The strong size quantization along the normal direction allows us to employ the adiabatic approximation and solve the Schr\"odinger equation for the effective Hamiltonian Eq.~(\ref{eq:effectivehamiltonianapprox}) considering the {\it ansatz} for the wavefunction $\chi(s,y,q_3)=\chi^{N}(s,q_3) \times \chi^{T} (s,y)$ where the normal wavefunction $\chi^N$ 
solves at fixed $s$  the one-dimensional Schr\"odinger equation for the ``fast'' normal quantum degrees of freedom 
\begin{equation}
 \left[-\dfrac{\hbar^2}{2 m^{\star}} \partial_{q_{3}}^2 + \widetilde{\gamma}(s) \kappa(s) q_3 + {\cal V}_{\lambda}(q_3) \right] \chi^{N}_i = E^{N}_i(s)  \chi^{N}_i. 
\label{eq:schfast}
\end{equation}
Here $i$ indicates the transversal subband index and   $\widetilde{\gamma}(s)=\gamma - \hbar^2 \left[\kappa(s)^2 + \partial_s^2 \kappa(s) / \kappa(s) \right] / (4 m^{\star})$ corresponds to the typical energy scale of the deformation potential locally renormalized by curvature effects. 
The latter term yields an attraction towards regions  under compressive strain (where the local curvature is higher) competing with the dominant  strain-induced attraction towards the tensile regions of the nanostructure [c.f. Fig.~\ref{fig:normalwaves}(a)]. 

The Hamiltonian for the slow tangential quantum motion can be  found by first integrating out the $q_3$ quantum degree of freedom and then performing the additional rescaling of the tangential wavefunction $\chi^T \rightarrow \chi^T \times  \sqrt{1 + 2 \kappa(s) \langle \chi^N | q_3 | \chi^N \rangle}$.
The final form of the dimensionally reduced tangential Hamiltonian is as follows
\begin{equation}
{\cal H}^T=-\dfrac{\hbar^2}{2 m^{\star}_R(s)} \partial_s^2 -\dfrac{\hbar^2}{2 m^{\star}} \partial_y^2 -\dfrac{\hbar^2}{8 m^{\star}} \kappa(s)^2 + E^N(s) + \ldots
\label{eq:schroslow2}
\end{equation}
where the $\ldots$ indicate the diagonal adiabatic corrections and we introduced the locally renormalized effective mass $m^{\star}_R(s)=m^{\star} / \left[1 + 2 \kappa(s) \langle \chi^N | q_3 | \chi^N \rangle \right]$. Fig.~\ref{fig:normalwaves}(b) shows the deviation of the locally renormalized effective mass from its bare value for different values of the expansion parameter $\kappa(s) \delta$. The strain-induced localization of the normal wavefunction in the tensile regions of the nanostructure leads to an heavier effective local mass 
 enhanced at the points of maximum curvature. This is at odds with the lighter local effective mass one would find in the absence of strain effects. 
Even for values of the local energy scale  $\widetilde{\gamma}(s)$ much larger than the characteristic energy of the normal quantum well energy $\hbar^2 / (2 m^{\star} \delta^2)$, this local renormalization of the effective mass is so small that for all practical purposes the use of the bare effective mass in Eq.~(\ref{eq:schroslow2}) is justified. 
More rigorously one can show the reliability of this approximation in the regime $\widetilde{\gamma}(s)\, \kappa(s) \, \delta \leq \hbar^2 / (2 m^{\star} \delta^2)$ where the strain-induced linear potential appearing in the fast Schr\"odinger equation Eq.~(\ref{eq:schfast}) can be treated perturbatively. Since such a condition is typically satisfied in conventional semiconducting nanostructures with a total thickness in the nanometer scale, we will limit ourselves to this regime from here onwards. 

In Fig.~\ref{fig:normalwaves}(c) we show the behavior of the adiabatic potentials in the first two transversal subband measured from the normal quantum well levels $E_i^{0 \, N}$'s. The characteristic quadratic dependence on the local energy scale $\widetilde{\gamma}(s)$ is well reproduced by the analytical formula $E_i^N(s)=E_i^{0 \, N}+ 2 \, m^{\star}\, \delta^4 \, \widetilde{\gamma}(s)^2 / \hbar^2 \times f_i \, \kappa(s)^2$ with the 
numerical constants that can be calculated using second-order perturbation theory as  $f_1 \sim -10^{-3}$, $f_2 \sim 3 \cdot 10^{-4}$, etc. From this it also follows that the distance among the potential energy surfaces $E_2^N(s)-E_1^N(s)>E_2^{0\, N}-E_{1}^{0 \, N} $  thereby guaranteeing the reliability of the adiabatic approximation in the regime $\kappa(s) \delta \ll 1$. As a result, we then find an effective Hamiltonian for the electronic motion along the curved nanostructure which in the first transversal subband reads
\begin{widetext}
\begin{equation}
 {\cal H}^{T}=-\dfrac{\hbar^2}{2 m^{\star}} \left[ \partial_y^2 + \partial_s^2\right] -\dfrac{\hbar^2  \kappa(s)^2}{8 m^{\star}} v_R +  |f_1| \, \gamma \,\delta^4 \, \kappa(s)
\left[\kappa(s)^3+ \partial_s^2 \kappa(s) \right] - \dfrac{\hbar^2 \delta^4 |f_1|}{8 m^{\star}}  \left[\kappa(s)^3+ \partial_s^2 \kappa(s) \right] ^2, 
\label{eq:finalhamiltonian}
\end{equation}
\end{widetext}
where we left out the constant energy term $\hbar^2 \pi^2 / (2 m^{\star} \delta^2)$ and we neglected the diagonal adiabatic corrections.  
The second term in Eq.~\ref{eq:finalhamiltonian} corresponds to a  geometric potential whose strength is renormalized by strain effects as $v_R=1+4 |f_1|  \left(2 m^{\star} \delta^2 \gamma / \hbar^2  \right)^2$. Remarkably we find this renormalization to become extremely large in case of the natural hierarchy of energy scales 
\begin{equation} 
 \gamma \gg \dfrac{\hbar^2}{2 m^{\star} \delta^2} \gg \dfrac{\hbar^2}{2 m^{\star} R(s)^2}.
\label{eq:hierarchy}
\end{equation}
Considering for instance a  value of $\gamma \sim 10$ eV,  a characteristic quantum well energy  $\hbar^2 / (2 m^{\star} \delta^2) \sim 5$ meV and the typical tangential kinetic energy $\hbar^2 /  [2 m^{\star} R(s)^2] \sim 1 \mu$eV, we find an enhancement of the curvature-induced potential by $\sim10^4$.  
As we show below, this gigantic renormalization of the  geometric potential has profound consequences on the electronic properties of low-dimensional nanostructures with curved geometry. 
It is worth noting that in absence of strain effects ($\gamma \rightarrow 0$), Eq.~(\ref{eq:finalhamiltonian}) corresponds to the effective tangential Hamiltonian introduced by Da Costa augmented with an higher-order curvature induced geometric potential $\propto \kappa(s)^6$ arising as a consequence of the finite thickness of the thin-film nanostructure. 

\begin{figure}[tbp]
\includegraphics[width=.9\columnwidth]{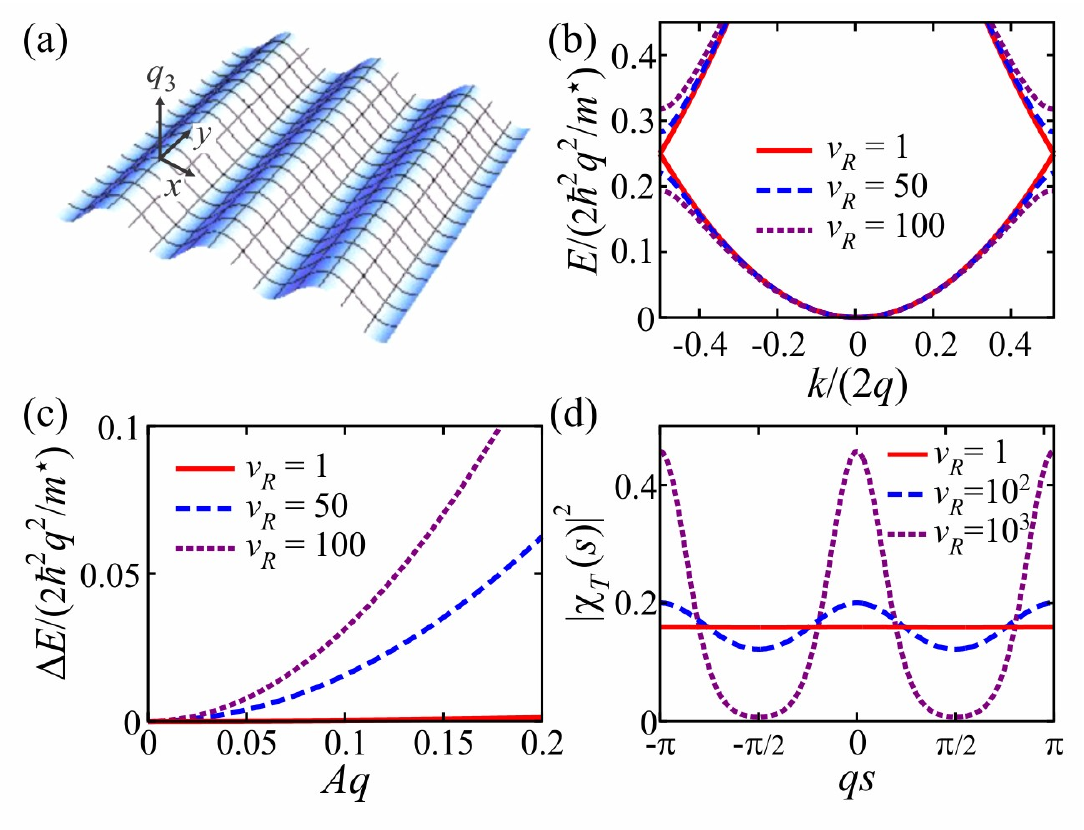}
\caption{(a) Sketch of the stress-free surface of a nanocorrugated thin film. (b) Bandstructure of a nanocorrugated film with $A q \equiv 0.2$ and different values of the geometric potential renormalization factor $v_R$. We show only the first two bands close to the bottom of the conduction band. (c)  Behavior of the topological bandgap as a function of the dimensionless parameter $A \, q$ for different values of $v_R$.  (d) Tangential ground state density probability for $A q \equiv 0.2$ and different values of the renormalization strength of the geometric potential.}
\label{fig:bandstructure2d}
\end{figure}

We now use this theoretical framework to analyze the influence of the strain-induced geometric potential on the electronic states of a nanocorrugated thin film with period $2 \pi / q$ and total thickness $\delta$ [c.f. Fig.~\ref{fig:bandstructure2d}(a)]. The stress-free surface can be parametrized in the Monge gauge as  $q_3(x)=A \, \cos{q \, x}$ where $A$ is the amplitude of the corrugation.  In the shallow deformation limit $A q \ll 1$, we can express the arclength of the layer $s \simeq x$ whereas the local curvature of the stress free surface $\kappa(s) \simeq -A \, q^2 \cos{(q \, s)}$.  Thus, the problem reduces to the "flat" motion of free electrons embedded in a curvature induced periodic potential. 
It is straight-forward to obtain the energy spectrum of the "slow'' tangential motion of Hamiltonian Eq.~(\ref{eq:finalhamiltonian}) in the first transversal subband  for a thin film thickness much smaller than the corrugation wavelength as in this case the last two terms  in Eq.~(\ref{eq:finalhamiltonian}) can be neglected.
Fig.~\ref{fig:bandstructure2d}(b) shows the behavior of the first two bands for different values of $v_R$. The zero of the energy has been chosen as the bottom the conduction band.  The qualitative behavior of the bandstructure is maintained when strain effects are taken into account, but now with curvature induced gaps  \cite{ono09} at momenta $k= \pm q$. The gaps increase quadratically in magnitude with the energy scale of the deformation potential $\gamma$. Indeed we find  $\Delta E \sim v_R \,\hbar^2  A^2 \, q^4 / (16 m^{\star})$, in agreement with the numerical analysis [c.f. Fig.~\ref{fig:bandstructure2d}(c)].  In Fig.~\ref{fig:bandstructure2d}(d)  we show the ground state density probability for different values of $v_{R}$. By increasing strain effects, one finds a continuous crossover from extended-like states to electronic states localized precisely at the points of maximum and minimum curvature, which is in accordance with the results of a purely numerical approach~\cite{osa05}.

In conclusion, by employing a method of adiabatic separation of fast and slow quantum degrees of freedom, we have derived a dimensionally reduced Schr\"odinger equation in strain-driven nanostructures. The strain effects render an often gigantic renormalization of the curvature induced geometric potential which has profound consequences upon the electronic properties of these materials. Applying our theoretical framework to the case of nanocorrugated thin films we find an enhanced electron localization and the opening of substantial band gaps on an experimentally relevant energy scale. It can also be applied to, for instance, 2D  nanotubes rolled-up in the shape of an Archimedean spiral, where the effect of the geometric quantum potential leads to bound states whose number coincides with the winding number \cite{ort10}. The inclusion of  strain effects will lead to a proliferation of such bound states, strongly affecting the electronic and transport properties of these rolled-up  nanostructures. 

The authors wish to thank P. Cendula and V. M. Fomin for fruitful discussions.

\end{document}